%% file: main.tex
\documentclass[conference]{IEEEtran}
\IEEEoverridecommandlockouts
\usepackage{cite}
\usepackage{amsmath,amssymb,amsfonts}
\usepackage{algorithmic}
\usepackage{graphicx}
\usepackage{textcomp}
\usepackage{xcolor}
\usepackage{enumitem}
\usepackage{float} 
\usepackage{booktabs}
\usepackage{fancyhdr}
\usepackage{balance}
\usepackage[T1]{fontenc}
\usepackage{aecompl}
\usepackage{makecell}
\usepackage[version=4]{mhchem}
\usepackage[hyphens]{url}
\usepackage{breakurl,bm}
\usepackage{multirow}
\usepackage{multicol}
\usepackage{booktabs}
\usepackage{comment}
\usepackage{pifont}
\usepackage{threeparttable}
\usepackage[switch]{lineno}

\def\BibTeX{{\rm B\kern-.05em{\sc i\kern-.025em b}\kern-.08em
    T\kern-.1667em\lower.7ex\hbox{E}\kern-.125emX}}

\begin{document}
% \linenumbers

\title{ 

Breaking the Training Barrier of Billion-Parameter Universal Machine Learning Interatomic Potentials %on Exascale Supercomputers

\thanks{\# These authors contributed equally to this work.}
\thanks{Corresponding authors: Weile Jia (jiaweile@ict.ac.cn)}

}

\author{
\IEEEauthorblockN{
Yuanchang Zhou\IEEEauthorrefmark{1}\IEEEauthorrefmark{2}\textsuperscript{\#}, 
Hongyu Wang\IEEEauthorrefmark{1}\IEEEauthorrefmark{2}\textsuperscript{\#}, 
Yiming Du\IEEEauthorrefmark{1}\IEEEauthorrefmark{2}\textsuperscript{\#}, 
Yan Wang\IEEEauthorrefmark{1}\IEEEauthorrefmark{2}\textsuperscript{\#}, 
Mingzhen Li\IEEEauthorrefmark{1}\IEEEauthorrefmark{2}\textsuperscript{\#}, 
Siyu Hu\IEEEauthorrefmark{1}\IEEEauthorrefmark{2}\textsuperscript{\#}, 
Xiangyu Zhang\IEEEauthorrefmark{1}\IEEEauthorrefmark{2},\\
Weijian Liu\IEEEauthorrefmark{1}\IEEEauthorrefmark{2},
Chen Wang\IEEEauthorrefmark{1}\IEEEauthorrefmark{2},
Zhuoqiang Guo\IEEEauthorrefmark{3},
Long Wang\IEEEauthorrefmark{3}, 
Jingde Bu\IEEEauthorrefmark{3}, 
Yutong Lu\IEEEauthorrefmark{4},
Guangming Tan\IEEEauthorrefmark{1} and
Weile Jia\IEEEauthorrefmark{1}
}
\IEEEauthorblockA{\IEEEauthorrefmark{1} State Key Lab of Processors, Institute of Computing Technology, Chinese Academy of Sciences
}
\IEEEauthorblockA{\IEEEauthorrefmark{2} 
University of Chinese Academy of Sciences
\IEEEauthorrefmark{3} 
Independent Researcher
\IEEEauthorrefmark{4} 
Sun Yat-Sen University
}
}

\newcommand{\framework}[0]{Janus}
\newcommand{\ZQ}[1]{\textcolor{magenta}{[ZQ:#1]}}
\newcommand{\REV}[1]{{\color{red}#1}}
\newcommand{\REQ}[1]{{\color{blue}(#1)}}
\newcommand{\WL}[1]{\textcolor{cyan}{[Weile:#1]}}
\newcommand{\zyc}[1]{\textcolor{purple}{[Yuanchang:#1]}}
\newcommand{\why}[1]{\textcolor{brown}{[Hongyu:#1]}}
\definecolor{mydarkgreen}{RGB}{0,100,0}
\newcommand{\wy}[1]{\textcolor{mydarkgreen}{[Wangyan:#1]}}
\newcommand{\zxy}[1]{\textcolor{orange}{[Xiangyu:#1]}}

\newcommand{\todo}[1]{{\color{blue}{[todo:#1]}}}
\maketitle
\thispagestyle{plain}

\begin{abstract}
Universal Machine Learning Interatomic Potentials (uMLIPs), pre-trained on massively diverse datasets encompassing inorganic materials and organic molecules across the entire periodic table, serve as foundational models for quantum-accurate physical simulations. However, uMLIP training requires second-order derivatives, which lack corresponding parallel training frameworks; moreover, scaling to the billion-parameter regime causes explosive growth in computation and communication overhead, making its training a tremendous challenge. We introduce MatRIS-MoE, a billion-parameter Mixture-of-Experts model built upon invariant architecture, and {Janus}, a pioneering high-dimensional distributed training framework for uMLIPs with hardware-aware optimizations. Deployed across two Exascale supercomputers, our code attains a peak performance of 1.2/1.0 EFLOPS (24\%/{35.5\%} of theoretical peak) in single precision at over 90\% parallel efficiency, compressing the training of billion-parameter uMLIPs from weeks to hours. This work establishes a new high-water mark for AI-for-Science (AI4S) foundation models at Exascale and provides essential infrastructure for rapid scientific discovery.

\end{abstract}


\section{Justification For Prize}

Record training of billion-parameter uMLIPs with 473 million atomic configurations. Single-precision performance reaches 1.2/1.0 EFLOPS, maintaining 90\%--93\% parallel efficiency on two Exascale systems. For the 11.5B model, normalized throughput is 653--3201$\times$ compared to the current state-of-the-art, establishing a new performance high watermark for training atomic AI4S models on Exascale systems.

\section{Performance Attributes}\label{sec:performance_attributes}

{
\begin{center}
\begin{tabular}{l | l} 
 \toprule
 Performance attribute & Our submission  \\
 \midrule
 Category of achievement  & Scalability, Throughput \\ 
 Type of method used  &  {Other: training of uMLIP} \\
 Results reported on basis of & Whole application with IO \\
 Precision reported & Single precision \\
 System scale & Measured on full system \\
 Measurements  & Timers, FLOP count\\
\bottomrule
\end{tabular}
\end{center}
}
\vskip 1em

\section{Overview Of The Problem\label{sec:overview}}

%\REQ{description of the problem and its importance, in terms understandable to a non-specialist (1 p max)}
\subsection{Molecular Dynamics and MLIPs}

Molecular dynamics (MD) simulations with quantum accuracy are an indispensable workhorse for modern scientific discovery, enabling breakthroughs in advanced materials~\cite{li_electronic_2007}, clean energy~\cite{ceder_identification_1998}, and pharmaceuticals~\cite{doi:10.1073/pnas.1516247112} by unveiling atomic-level mechanisms. Driven by AI for Science (AI4S), the field is now shifting from computationally expensive Kohn-Sham Density Functional Theory (KS-DFT) to highly efficient surrogate modeling.
Recent advancements in specialized MLIPs~\cite{fan2021neuroevolution, Batzner_2022}, such as DeePMD~\cite{10.5555/3433701.3433707}, have addressed the computational bottlenecks by training on \textit{ab initio} data to reach unprecedented spatio-temporal scales. Nevertheless, such specialized MLIPs often suffer from poor generalization and an inability to represent the elemental diversity of multi-component physical systems. This limitation stems from their restricted model capacity, which is typically limited to tens of thousands of parameters~\cite{fan2021neuroevolution, Batzner_2022}.

The growing complexity of scientific problems—such as high-throughput screening of solid-state electrolytes and iron-based catalysts—has driven the transition to universal MLIPs~\cite{wood2025uma,kim_optimizing_2026}. These challenges involve heterogeneous interfaces and intricate surface reactions governed by high-dimensional potential energy surfaces (PES) that span diverse chemical elements and chemical domains. To capture such complexity while preserving physical symmetries, the state-of-the-art uMLIPs are predominantly formulated as Graph Neural Networks (GNNs)~\cite{deng_2023_chgnet,zhou2026matris,wood2025uma}. For each atom $i$, the local environment within a cutoff radius is represented as a graph, with atoms as nodes and bonds as edges. Iterative message passing enables local representations to incorporate information from increasingly distant atoms, capturing multi-body correlations and effective long-range interactions while maintaining linear scaling with system size.

Driven by the explosive growth of large-scale open datasets~\cite{levine2026openpolymers2026opoly26, gharakhanyan2025openmolecularcrystals2025, sriram2025opendac2025dataset, sahoo2025open, chanussot2021open, kaplan2025foundationalpotentialenergysurface, kim_optimizing_2026}, such as OMat24~\cite{barroso2024open} and OMol25~\cite{levine2026openmolecules2025omol25}, uMLIPs have rapidly scaled from millions to the billion-parameter regime (Table~\ref{table:sota}). Consequently, the primary computational bottleneck has shifted entirely to the training phase, making the optimization of billion-parameter uMLIPs a formidable HPC challenge. Three fundamental factors exacerbate this difficulty: \textit{a). Force-matching via second-order derivatives:} Unlike the first-order optimizations used in Large Language Models (LLMs), uMLIP training requires strict force-matching ($F_i=-\partial E/\partial X_i$), necessitating double-backward (second-order automatic differentiation). This doubles both the computational costs and memory footprints from exploding intermediate activations. \textit{b). High-precision (FP32) requirement:} While LLMs routinely exploit low-precision formats (e.g., FP16 or FP8), uMLIPs must use full single-precision (FP32) arithmetic to preserve the quantum-accurate fidelity essential for stable molecular dynamics. This constraint doubles the memory footprint and severely limits tensor-core utilization on many-core architectures. \textit{c). Extreme edge-token throughput demand:}  Message passing along edges dominates the computational cost, making the edges the fundamental ``tokens'' of uMLIP training. The 473 million configurations used in this work yield approximately 3.6 trillion interacting edges, imposing unprecedented demands on system-level throughput.

\subsection{MatRIS}

\begin{figure}
    \centering
    \includegraphics[width=1.0\linewidth]{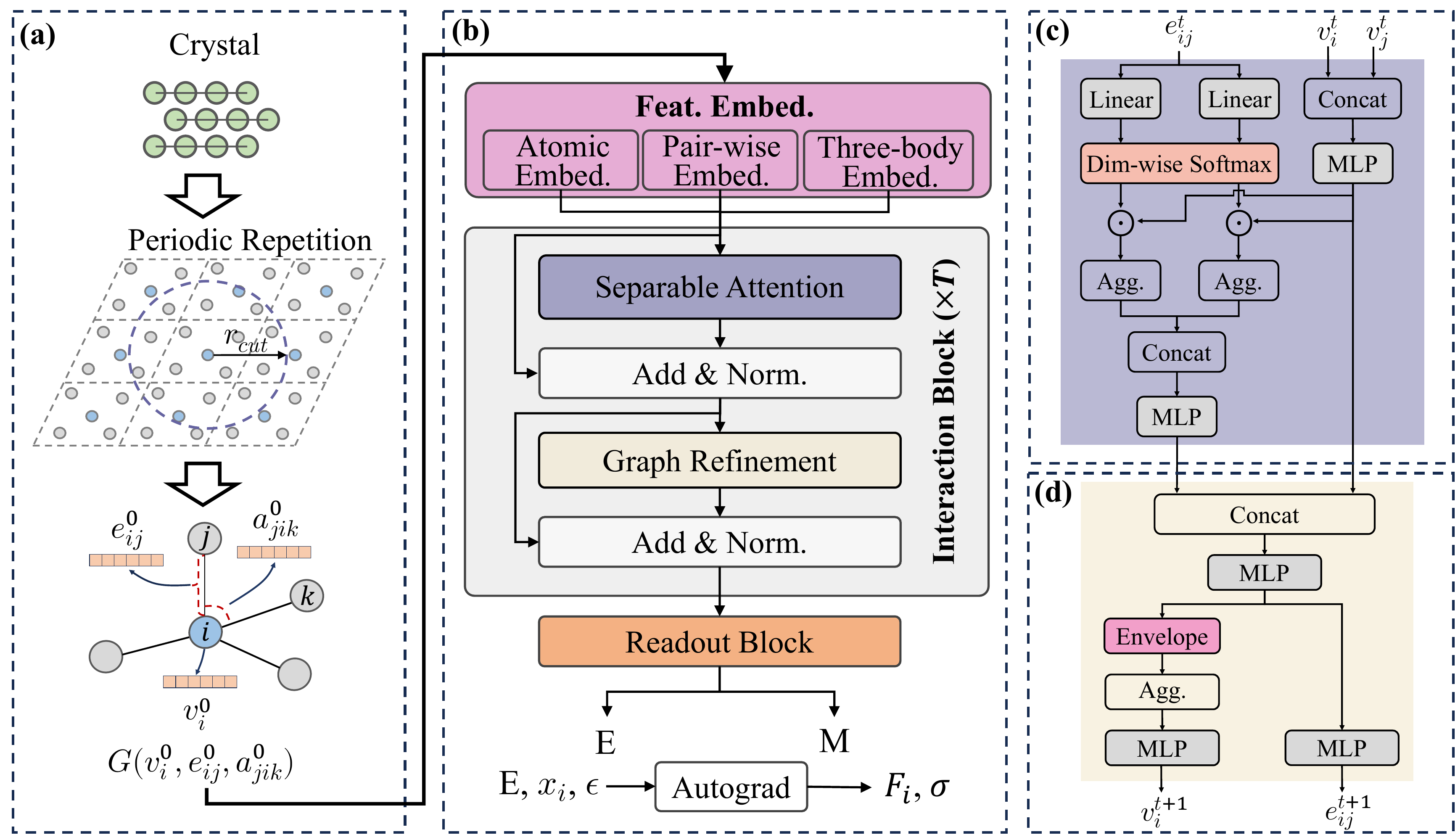}
    \caption{
    Overview of the MatRIS architecture.
    (a) Graph construction under periodic repetition condition.
    (b) Overall workflow, including feature embedding, interaction blocks, and readout.
    (c) Graph separable attention for invariant message passing.
    (d) Graph Refinement module.
    }
    \label{fig:matris_v1}
\end{figure}

Materials Representation and Interaction Simulation (MatRIS) is an invariant, attention-based uMLIP designed for highly efficient atomistic modeling. As illustrated in Fig.~\ref{fig:matris_v1}(a), \ding{202} an atomistic system is represented as $G(Z,X,L)$, where $Z$, $X$, and $L$ denote atomic numbers, Cartesian coordinates, and lattice vectors, respectively. Under periodic boundary conditions, the structure is first periodically repeated and then converted by \textit{Graph Generation Module} into the graph structure $G(v^0_i,e^0_{ij}, a^0_{ijk})$.
% atom graph $G^a=(V^a,E^a)$, and a line graph $G^l=(V^l,E^l)$. In $G^a$, atoms are represented by nodes, and bonds are represented by edges. 
% In $G^l$, bonds in $E^a$ are converted to nodes, and angles (two bonds share an atom) are represented by edges.
Fig.~\ref{fig:matris_v1}(b) shows the architecture of MatRIS. \ding{203} In the \textit{Embedding Module}, atomic numbers, pairwise distances, and angles are further embedded as initial node, edge, and angular features. \ding{204} MatRIS adopts a message-passing mechanism with $L$ \textit{Interaction Blocks} (IBs). In $\mathrm{IB^t}$, $\{ v^t_i, e^t_{ij}, a^t_{ijk} \}$ are iteratively transformed to $ \{ v^{t+1}_i, e^{t+1}_{ij},  a^{t+1}_{ijk}\} $, where $t=1, \dots, T-1$, $T$ is the number of IBs. Each IB consists of a Graph Separable Attention module (Fig.~\ref{fig:matris_v1}(c)) and a Refinement module (Fig.~\ref{fig:matris_v1}(d)). In the Separable Attention module, two linear projections of $e_{ij}$ are normalized by dim-wise softmax over the target and source neighborhoods to obtain the score of target and source atoms, and then the weighted messages are aggregated separately and concatenated to update the node feature $v_i^{'}$. By replacing dense pairwise attention with source- and target-aware neighborhood aggregation, separable attention reduces the attention complexity from $O(N^2)$ to $O(N)$. Finally, \ding{205} the \textit{Readout Block} predicts total energy $E$ and magnetic moments $M$. Forces and stress are obtained by automatic differentiation, $F_i=-\frac{\partial E}{\partial X_i}$ and $\sigma=\frac{1}{\mathcal{V}}\frac{\partial E}{\partial \epsilon}$, where $\mathcal{V}$ denotes volume.

Extensive benchmarking shows that the 10-million-parameter MatRIS establishes a new Pareto frontier in both accuracy and training efficiency. On the Matbench-Discovery benchmark, it achieves a record F1 score of 0.847 and an RMSD of 0.0717 while delivering a $13\times$ training speedup over state-of-the-art uMLIPs~\cite{zhou2026matris}. The model further demonstrates strong zero-shot generalization across diverse downstream tasks~\cite{zhou2026matris}. In this work, we extend MatRIS to MatRIS-MoE by incorporating the Mixture-of-Experts (MoE) architecture to enable multi-task learning across heterogeneous chemical domains and datasets (Sec.~\ref{subsec: model inno}).

% \WL{I need a figure here, the figure is the architecture of MatRIS. }
% \zyc{The base architechture of MatRIS are described in Figure~\ref{fig:matris_v1}}
% \WL{can we add some notation on this figure to connect the text to it? will discuss it with Yuanchang later.}

\section{Current State Of The Art}

\begin{table*}[htb]
\vspace{-0.2cm}
\caption{ Comparison of state-of-the-art universal Machine Learning Interatomic Potentials (uMLIPs) published between 2022 and 2026.   }
\centering
\vspace{-0.2cm}
\label{table:sota}
\begin{threeparttable}
%\resizebox{\textwidth}{!}{%
\setlength{\tabcolsep}{5pt}
\begin{tabular}{l c c c c c c c c c c c} 
\toprule
Work & Year & Category & Multi-task & Dataset & \makecell[c]{\#Total \\ Params} & \makecell[c]{\#Active \\ Params} & Hardware & \makecell[c]{\#GPUs/ \\ \#Cores} & \makecell[c]{Peak} & \makecell[c]{FLOPS} &  \makecell[c]{Norm. Throughput \\ {[UMA=1]} }\\
\hline

\makecell[l]{M3GNet \cite{chen_universal_2022}} & 2022 & Invariant & No & 0.19M & 0.23M & 0.23M &  RTX3090 & 1 & -- & -- & -- \\
%\makecell[l]{DPA2 \cite{zhang_dpa-2_2024}} & 2022 & Equivariant & Yes & 4.05M & 7.68M & 7.68M & -- & -- & -- & -- & -- \\
\makecell[l]{CHGNet \cite{deng_2023_chgnet}} & 2023 & Invariant & No & 1.58M & 0.41M & 0.41M &  A100 & 1 & -- & -- & 0.0022 \\

\makecell[l]{GNoME \cite{merchant_scaling_2023}} & 2023 & Equivariant & No & 89.0M & 16.2M & 16.2M & TPUv3 & 4 & -- & -- & -- \\

\makecell[l]{eqV2 \cite{barroso2024open} }& 2024 & Equivariant & No & 113M & 86M & 86M &  A100 & 64 & -- & -- & 3.09 \\

%\makecell[l]{MatterSim \cite{yang2024mattersimdeeplearningatomistic} }& 2024 & Equivariant & Yes & 17.0M & 4.6M & 4.6M & -- & -- & -- & -- & -- \\

\makecell[l]{ORB \cite{rhodes2025orbv3atomisticsimulationscale} }& 2025 & Unconstrained & No & 133M & 25.5M & 25.5M &  A100 & 8 & -- & -- & -- \\

\makecell[l]{MatRIS-L \cite{zhou2026matris} }& 2026 & Invariant & No & 113M & 10.4M & 10.4M &  A800 & 128 & -- & -- & 0.565 \\

\makecell[l]{PET \cite{bigi2026pushinglimitsunconstrainedmachinelearned} }& 2026 & Unconstrained & No & 113M & 730M & 730M &  H100 & 512 & -- & -- & -- \\

\makecell[l]{UMA \cite{wood2025uma}}& 2025 & Equivariant & Yes & 459M & 1.4B & 50M &  H200 & 256 & -- & -- & 1.00 \\

\makecell[l]{MACE-mh \cite{batatia2025crosslearningelectronicstructure} }& 2025 & Equivariant & Yes & 117M & 6.4M & 6.4M &  H100 & 48 & -- & -- & -- \\

\makecell[l]{SevenNet-Omni \cite{kim_optimizing_2026} }& 2026 & Equivariant & Yes & 243M & 54.9M & 54.9M &  H200 & 8 & -- & -- & 0.706 \\

\hline
\makecell[l]{This work (M)} & 2026 & Invariant & Yes & 473M & 2.47B & 0.56B & GPGPU & 45K & 25.44\% & 750.9P & 653.4 \\

\makecell[l]{This work (L)} & 2026 & Invariant & Yes & 473M & 11.50B & 2.89B & GPGPU & 45K & 35.52\% & 1048P & 2795.9 \\

\makecell[l]{This work (M)} & 2026 & Invariant & Yes & 473M & 2.47B & 0.56B & ARMv9 & 12.4M & 17.53\% & 861.6P & 749.9 \\

\makecell[l]{This work (L)} & 2026 & Invariant & Yes & 473M & 11.50B & 2.89B & ARMv9 & 12.4M & 24.41\% & 1200P & 3201.8 \\
\hline

\end{tabular}

\begin{tablenotes}
\footnotesize
\item[*] This table details model capacity (total and active parameters), multi-task learning capability, dataset size, training hardware configuration, and core performance metrics. 
Peak performance (as a percentage of the theoretical peak) and FLOPS are reported for the training of the uMLIP models.
Normalized Throughput is calculated via $\text{\#Active Params} {\times} (\text{Dataset Size} {\times} \text{Epochs}) / \text{Training Days}$. The throughput of UMA is 1,050, which is normalized to 1.0. The symbol ``--'' denotes data not reported in the respective literature. This work reports the training of the medium (M, 2.47B) and large (L, 11.50B) variants of the MatRIS-MoE.
\end{tablenotes}
\end{threeparttable}
\end{table*}

\subsection{Scaling of the uMLIPs}
The ultimate goal of uMLIPs is to enable zero-shot, \textit{ab initio-level} accurate simulations across the entire periodic table, thereby unifying traditionally isolated domains such as organic molecules, crystalline solids, and catalytic interfaces. However, enforcing 3D spatial symmetry—a fundamental physical requirement—imposes significant architectural trade-offs. Equivariant models (e.g., eqV2~\cite{liao2024equiformerv}, MACE~\cite{Batatia2022mace}) natively satisfy these symmetries via high-degree irreducible representations; yet their reliance on Clebsch-Gordan tensor products introduces substantial computational complexity and memory overhead. Conversely, unconstrained models (e.g., ORB~\cite{rhodes2025orbv3atomisticsimulationscale}, PET~\cite{bigi2026pushinglimitsunconstrainedmachinelearned}) implicitly learn symmetries through extensive data augmentation, circumventing complex tensor operations. While this approach enhances inference efficiency, it introduces potential thermodynamic inconsistencies, such as non-conservative forces during long-timescale molecular dynamics simulations. Invariant models (e.g., CHGNet~\cite{deng_2023_chgnet}, DPA3~\cite{zhang2026dpa3}, and MatRIS~\cite{zhou2026matris}) ensure rotational and translational invariance by utilizing distances and angles. Although computationally efficient, invariant models have not yet been scaled to the billion-parameter regime using MoE architectures. Matching the fidelity of billion-parameter equivariant models in capturing complex, high-order multi-body interactions therefore requires a significantly larger parameter space.

To bridge this accuracy-efficiency gap and support diverse chemical domains, the community has rapidly increased both dataset size and model capacity, driving an exponential rise in computational demands. As detailed in Table \ref{table:sota}, the hardware configurations required to train representative uMLIPs have escalated from a single GPU to large-scale clusters. In 2023, training the lightweight CHGNet (0.41M parameters) required 8.3 days on a single A100 GPU. By 2024, training the 86M-parameter eqV2 model necessitated 6 days across a cluster of 64 A100 GPUs. In 2025, the 1.4B-parameter UMA model utilized 256 H200 GPUs for 21 continuous days. Even with recent architectural advancements, hardware requirements remain massive; for instance, the 730M-parameter PET model requires the parallel scheduling of 512 H100 GPUs.

Despite this rapid development, scaling uMLIPs into the billion-parameter regime confronts formidable barriers. First, current uMLIP training relies predominantly on pure data parallelism. As active parameters reach the billion scale, the memory footprint inherently exceeds the capacity of a single GPU. Moreover, state-of-the-art frameworks optimized for first-order LLMs are fundamentally incompatible with the second-order computational graphs required for uMLIP force-matching. This lack of native model parallelism effectively stalls the development of larger models. Second, a tenfold increase in active model parameters will dramatically extend training time on conventional clusters of only hundreds of GPUs, rendering brute-force scaling economically unaffordable. Finally, while algorithmic innovations have advanced rapidly, system-level hardware–software co-design has significantly lagged behind, resulting in poor utilization of modern HPC and severely limiting overall training efficiency.

\subsection{Scaling MatRIS to Billion-parameter Regime}

Building on the 10-million-parameter MatRIS—an invariant SOTA uMLIP originally trained with pure data parallelism on 128 A800 GPUs—we scale the architecture to the billion-parameter regime to create a true multi-task foundation model. This extension enables joint learning across heterogeneous domains (molecules, materials, catalysis, MOFs, and direct air capture) on Exascale systems. The scaling process has three major challenges:
(1) \textit{Architectural adaptation for multi-task learning}: Expanding the parameter space requires integrating a Mixture-of-Experts (MoE) architecture to handle heterogeneous chemical rules across tasks. In addition, although the original $O(N)$ separable attention is theoretically efficient, its fragmented and memory-bound operations result in low utilization of Tensor Cores on modern accelerators, necessitating a shift toward more compute-dense operators such as standard self-attention. 
(2) \textit{Lack of second-order parallel training framework}: As active parameters reach the billion scale, a single GPU can no longer accommodate the uMLIP model. Consequently, pure data parallelism is no longer applicable, necessitating the development of a second-order parallel training framework.
(3) \textit{Bottlenecks on large-scale platforms}: Training billion-parameter uMLIPs on large-scale many-core supercomputers (e.g., GPU-based and ARM-based Exascale systems) significantly increases cross-node communication volume and exposes the inefficiency of second-order operators, limiting overall parallel scalability.

\section{Innovations Realized}

\input{3-model}
\input{012-system}

\section{How Performance Was Measured}

\subsection{Physical System Used to Measure Performance}

To achieve universal generalizability, we construct a large multi-domain dataset comprising over 473 million atomic configurations spanning isolated molecules, periodic crystals, catalytic surfaces, molecular crystals, and metal-organic frameworks (MOFs)~\cite{barroso2024open,levine2026openmolecules2025omol25, levine2026openpolymers2026opoly26, gharakhanyan2025openmolecularcrystals2025, sriram2025opendac2025dataset, sahoo2025open, chanussot2021open, kaplan2025foundationalpotentialenergysurface, kim_optimizing_2026}. 
%\REV{[Add citation OPoly26, OMC25, ODAC25, OC25, OC20, MatPES, SevenNet-Omni]}.
Ground-truth labels for system energy, atomic forces, and virial stresses are obtained from high-fidelity DFT calculations, with each configuration represented as a unique spatial graph. The MatRIS-MoE architecture exhibits strict linear scaling, $O(N)$, where $N$ denotes the total number of edges in the spatial graphs. Notably, our 473M dataset contains approximately 3.6 trillion edges, posing significant challenges for system throughput.

We evaluate three MatRIS-MoE variants: small (S), medium (M), and large (L). 
Due to time and computational constraints, only the small model is fully trained; the (M) and (L) variants are used exclusively for weak and strong scaling studies. 
All three configurations share the same backbone architecture, consisting of six interaction layers with a MoE routing mechanism and multi-head self-attention (Fig.~\ref{fig:overview}(a)). Detailed hyperparameters are summarized in Table~\ref{tab:model_params}.

\begin{table}[htbp]
\vspace{-0.2cm}
\centering
\caption{Configurations of MatRIS-MoE (S), (M) and (L) models.}
\vspace{-0.2cm}
\label{tab:model_params}
\begin{tabular*}{\columnwidth}{@{\extracolsep{\fill}} l c c c @{}}
\toprule 
{Hyperparameter} & {(S)} & {(M)} & { (L)} \\
\midrule 
Total parameters & 1.09 B & 2.47 B & 11.5 B \\
Active parameters & 0.19 B & 0.56 B & 2.89 B \\
Number of experts & 72 & 40 & 72 \\
Top-$k$ routing & 4 & 8 & 16 \\
Number of layers & 6 & 6 & 6 \\
Node feature dimension & 2560 & 1536 & 1920 \\
Edge feature dimension & 384 & 1536 & 1920 \\
MLP hidden dimension & 2560 & 1536 & 1920 \\
Number of attention heads & 8 & 4 & 4 \\
Attention dimension & 384 & 512 & 512 \\
\bottomrule 
\vspace{-0.2cm}
\end{tabular*}
\end{table}

\begin{figure}[h]
    \centering
    \includegraphics[width=\linewidth]{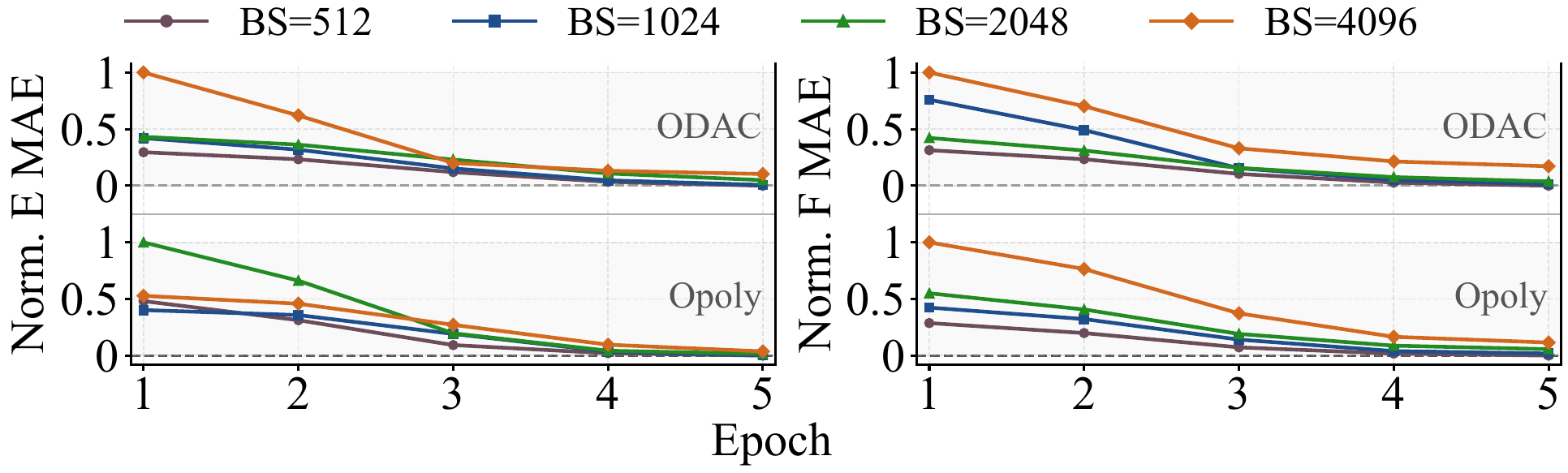}
    \caption{Convergence behavior of MatRIS-MoE under different batch sizes on the ODAC25\cite{sriram2025opendac2025dataset} and Opoly26\cite{levine2026openpolymers2026opoly26} validation sets.}
    \label{fig:batch_loss_test}
\end{figure}

Figure~\ref{fig:batch_loss_test} examines the convergence behavior under varying batch sizes using a 1\% sample of the entire dataset.
The left and right panels present the energy and force learning curves of MatRIS-MoE, respectively. By scaling the learning rate by the square root of the batch size multiplier, we successfully stabilize the optimization at extreme scales. 
Quantitatively, increasing the batch size from 512 to 4,096 (4K) results in less than a 10\% increase in the final energy loss.
Notably, in our Exascale evaluation, the maximum global batch sizes reach 4K and 15K on the CNIS and LineShine systems, respectively.

\subsection{HPC Systems and Software}

All performance measurements are conducted on two distinct Exascale supercomputers: the China New-generation Intelligent Supercomputer (CNIS) and the LineShine system.
The CNIS is a heterogeneous Exascale supercomputer consisting of $5,632$ computing nodes, denoted the full machine. 
Each node is equipped with two 64-bit CISC-based server processors and eight SIMT-based GPGPU accelerators. The host processor runs at 2.4 GHz with 64 cores in a NUMA architecture, connected to 8-channel DDR5-6400 memory and PCIe Gen5 interfaces, delivering 64 GB/s host-to-device bandwidth. Each GPGPU provides 32.7 TFLOPS (FP64), 65.5 TFLOPS (FP32), and 470 TFLOPS (FP16) peak performance, with 64 GB HBM (1.8 TB/s bandwidth), 320 SIMD units, 768 KB registers, 64 KB LDS, and 8 MB L2 cache. The accelerators are interconnected via high-speed chip-to-chip links, while nodes are connected through a proprietary InfiniBand-like RDMA network with a three-layer Clos dual-plane topology, providing $4\times 400$ Gbps per node.

The LineShine supercomputer, developed by the National Supercomputing Center in Shenzhen (NSCC-SZ), is an exascale system consisting of 20,480 computing nodes, denoted the full machine.
Each node is equipped with two ARMv9-based LX2 processors. 
Each LX2 integrates two compute dies (304 cores total) and eight on-package HBM stacks (32 GB, 4 TB/s aggregate bandwidth). Each compute die contains 152 cores and 128 GB of off-package DDR memory organized into four NUMA domains. A dedicated SDMA engine handles data movement between DDR and HBM. The LX2 supports FP64/FP32/FP16/INT8 via SME and SVE units, delivering up to 60.3/120.6 TFLOPS in FP64/FP32. Nodes are interconnected via the LingQi high-speed network with a dual-plane multi-rail fat-tree topology, offering 1.6 Tb/s bandwidth per node. 

% software stacks
On CNIS, experiments run on Anolis OS 8.9 with a ROCm-compatible environment, including GCC 8.5.0, rocBLAS, and PyTorch 2.7.1. Inter-node communication uses OpenMPI 5.0.3 and an RCCL-compatible collective library, with a strict one-to-one mapping of 8 RCCL ranks per node, each bound to a single GPGPU. 
On LineShine, the environment consists of PyTorch 2.10.0, KML 25.2.1 BLAS, and OpenMPI 4.1.6rc4. To maximize the many-core efficiency of the LX2, we employ a hybrid MPI+OpenMP configuration with 16 MPI processes per node, each using multi-threading across 38 physical cores.

\subsection{Measurement Methodology}

The total floating-point operations (FLOPs) are measured in FP32 using the PyTorch Profiler. 
Since all compute-intensive operations are offloaded to accelerators, the profiled GPGPU FLOPs accurately represent the overall workload. Over 1,000 training iterations processing 30 million atomistic samples, the cumulative workload reaches {2974 EFLOPs} for the 11.50B-parameter MatRIS-MoE (L) and {1765 EFLOPs} for the 2.47B-parameter MatRIS-MoE (M). 
The following criteria are used to measure the performance of our code:

%The average number of edges processed in each iteration is 86.4$\times$ \#data processed $\times$

\begin{itemize}[leftmargin=*]

\item \textbf{Average throughput}, defined as $\frac{ \text{7,610} \times \text{global batch size}} {\text{average time per iteration}}$. Each atomistic configuration has $7610$ edges on average. 

\item \textbf{Peak performance}, defined as $\frac{ \text{total FLOPs}} {\text{MatRIS-MoE training loop time}}$. 

\item \textbf{Sustained performance}, defined as $\frac{ \text{total FLOPs}} {\text{total wall clock time}}$. The ``total wall clock time'' includes the whole application running time (including IO, {MPI initialization and finalization}).

\end{itemize}

\section{Performance Results} \label{sec:perfRes}

\subsection{Accuracy Results}
\begin{figure}[h]
    \centering
    \includegraphics[width=\linewidth]{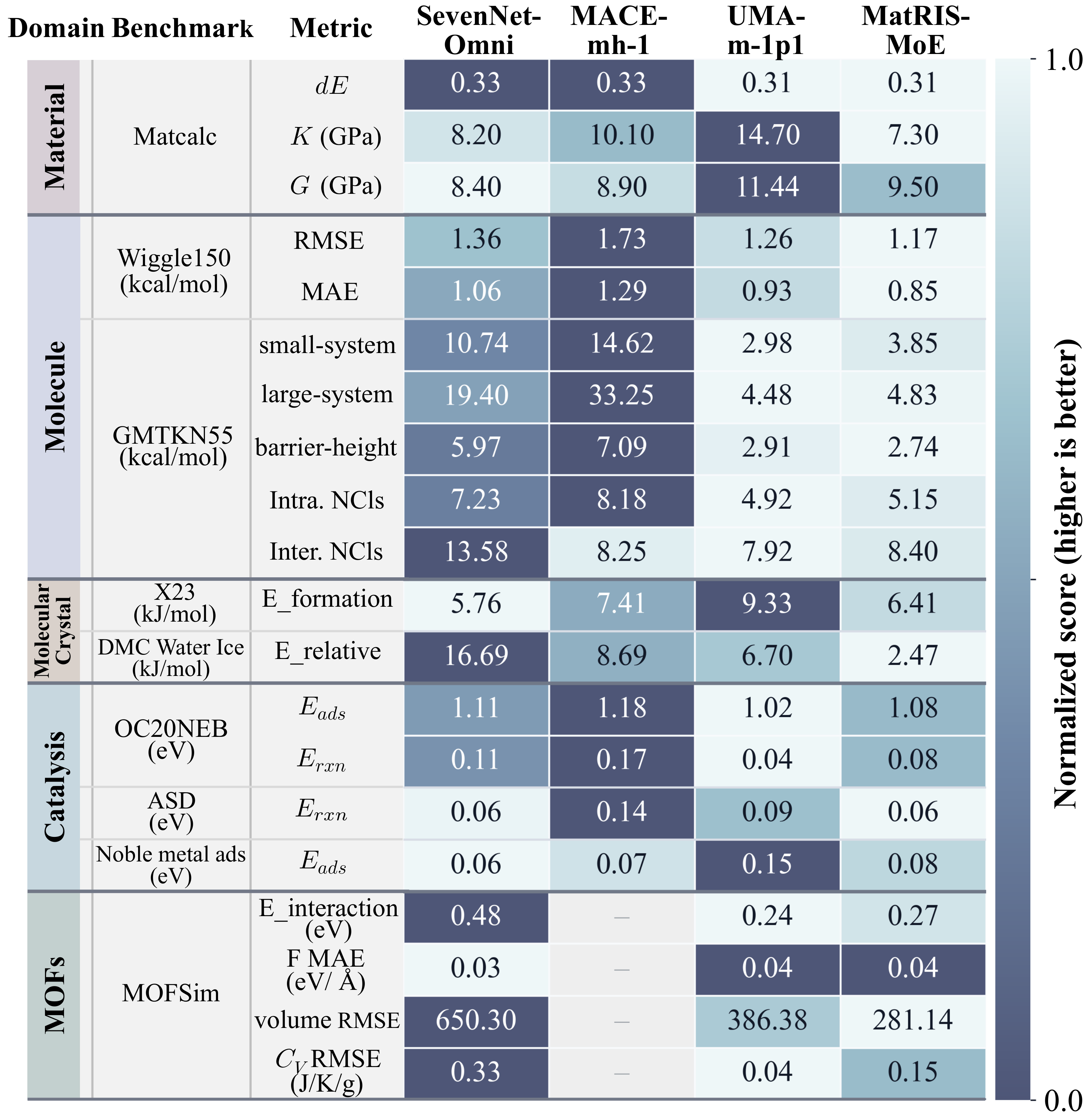}
    \caption{The out-of-the-box accuracy results of MatRIS-MoE on cross-domain benchmarks. The representative tasks span molecules, materials, catalysis, molecular crystals, and MOFs. ``-'' indicates that the corresponding results are not evaluated because the training dataset of MACE-mh-1  does not include MOF configurations. Lighter colors indicate higher accuracy.}
    \label{fig:res_benchmark}
\end{figure}

% Siyu(04-14)
Although MatRIS-MoE (S) has only about 1 billion parameters, it achieves state-of-the-art (SOTA) or near-SOTA accuracy across multiple benchmarks. 
These tasks comprise highly complex, real-world downstream applications and effectively demonstrate the generalization capacity of MatRIS-MoE. 
As shown in Fig.~\ref{fig:res_benchmark}, the evaluation spans a broad spectrum of systems, including materials (Matcalc~\cite{kaplan2025foundationalpotentialenergysurface}, reporting structural similarity $dE$, bulk moduli $K$ and shear moduli $G$), molecules (Wiggle150~\cite{doi:10.1021/acs.jctc.5c00015} with energy RMSE and MAE; GMTKN55~\cite{C7CP04913G} with small-system, large-system, barrier-height, and intra-/intermolecular NCI subsets), molecular crystals (X23~\cite{10.1063/1.4812819} formation energies and DMC Ice~\cite{10.1063/5.0102645} relative energies), catalysis (OC20NEB-OOD~\cite{wander2024cattsunamiacceleratingtransitionstate} with adsorption energies $E_{ads}$ and reaction energies $E_{rxn}$, ASD $E_{\mathrm{rxn}}$, and noble-metal adsorption~\cite{kim_optimizing_2026} with $E_{\mathrm{ads}}$), and MOFs (MOFSimBench~\cite{High_temper_phase_transition} with interaction-energy MAE, force MAE, volume RMSE and Heat capacity $C_V$). Lighter background colors indicate higher normalized accuracy.
These results highlight MatRIS-MoE’s strong generalization capability and representational power (e.g., achieving $\omega$B97M-V-level accuracy on Wiggle150\cite{doi:10.1021/acs.jctc.5c00015} while delivering an approximately three-orders-of-magnitude speedup), enabling a single model to deliver robust, out-of-the-box predictions across highly diverse chemical systems, including materials, molecules, catalytic surfaces, molecular crystals, and metal-organic frameworks.

\subsection{Performance Improvement}

In this section, we measure the reduction in per-step training time achieved by our system-level optimizations. 
All experiments are performed at 1/8 full-machine scale. 
For MatRIS-MoE, both the 2.47B-parameter (M) and 11.5B-parameter (L) models use the same parallel configuration, with an FS-3D unit size of 8 and a GP-replica size of 8, yielding 128 and 64 atoms per FS-3D unit, respectively.
Both configurations scale to the full system through the DP-replica dimension.
Table~\ref{tab:perf_improvement} summarizes the per-step training time before and after optimization, together with the resulting speedup. 

\begin{table}[htbp]
\centering
\vspace{-0.2cm}
\caption{Overall improvement on CNIS and LineShine.}
\label{tab:perf_improvement}
\vspace{-0.2cm}
\begin{tabular}{lcccc}
\toprule
{Machine} & {Model} & {Baseline} & {Optimized} & {Speedup} \\
\midrule
\multirow{2}{*}{CNIS}  & MatRIS-MoE (M)  & 5.96s & 2.21s & 2.7$\times$ \\
                           & MatRIS-MoE (L)  & 7.71s & 2.66s & 2.9$\times$ \\
\midrule
\multirow{2}{*}{LineShine} & MatRIS-MoE (M)  & 33.1s & 8.08s & 4.1$\times$ \\
                           & MatRIS-MoE (L)  & 40.7s & 8.14s & 5.0$\times$ \\
\bottomrule
\vspace{-0.2cm}
\end{tabular}
\end{table}

Although MatRIS-MoE (M) has fewer parameters than MatRIS-MoE (L) (2.47B vs. 11.5B), their per-step training times remain comparable. This is because MatRIS-MoE (M) processes twice as many atoms per FS-3D unit (128 vs. 64), resulting in a higher per-rank computational workload. Consequently, the theoretical per-step FLOP count of MatRIS-MoE (M) reaches 59.36\% of that of MatRIS-MoE (L).

The observed performance gains come from three system-level optimizations. 
First, we apply an asynchronous optimizer that overlaps gradient synchronization with parameter updates, thereby reducing synchronization stalls during double backward. 
Second, we apply the atom-type-aware compression that reduces the MoE routing all-to-all communication volume by 50\% without sacrificing numerical precision. 
% Third, we apply hierarchical collective operations tailored to the distinct network topologies of CNIS (dual-plane three-layer Clos) and LineShine (dual-plane multi-rail fat-tree), thereby maximizing inter-node bandwidth utilization.
Third, we design several high-performance kernels including graph operators, attention, and MoE dispatch/combine, thereby improving single-accelerator performance.

Together with the above communication optimizations, the up to 5.0$\times$ speedup on LineShine is further enabled by a platform-specific optimization: SDMA-based data movement between DDR and on-package HBM.
This addresses the limited HBM utilization of native PyTorch operators on the LX2 processor, allowing operators to achieve up to 1.4$\times$ higher memory bandwidth. % and the memory-bound kernels in the interaction blocks yields substantial acceleration.

% avg throughput
% CNIS - weak - large
% 1626919.294
% 3207308.697
% 6268845.425
% 11188649.37

% CNIS - weak - small
% 3920728.882
% 7741871.25
% 15146559.82
% 27001014.28

% CNIS - strong - large
% 1626919.294
% 1310646.184
% 1019590.322
% 804281.1109

% CNIS - strong - small
% 3920728.882
% 3111490.441
% 2374001.338
% 1864208.565

% lineshine avg throughput
% lineshine - weak - large
% 1772780.695
% 3476187.07
% 6710097.67
% 12809701.33

% lineshine - weak - small
% 4243240.434
% 8339046.242
% 17152760.06
% 30982847.1

% lineshine - strong - large
% 1772780.695
% 1386952.747
% 1060626.62
% 897196.7319

% lineshine - strong - small
% 4243240.434
% 3239965.789
% 2498297.704
% 2118473.746

\begin{figure}[htbp]
    \centering
    \includegraphics[width=\linewidth]{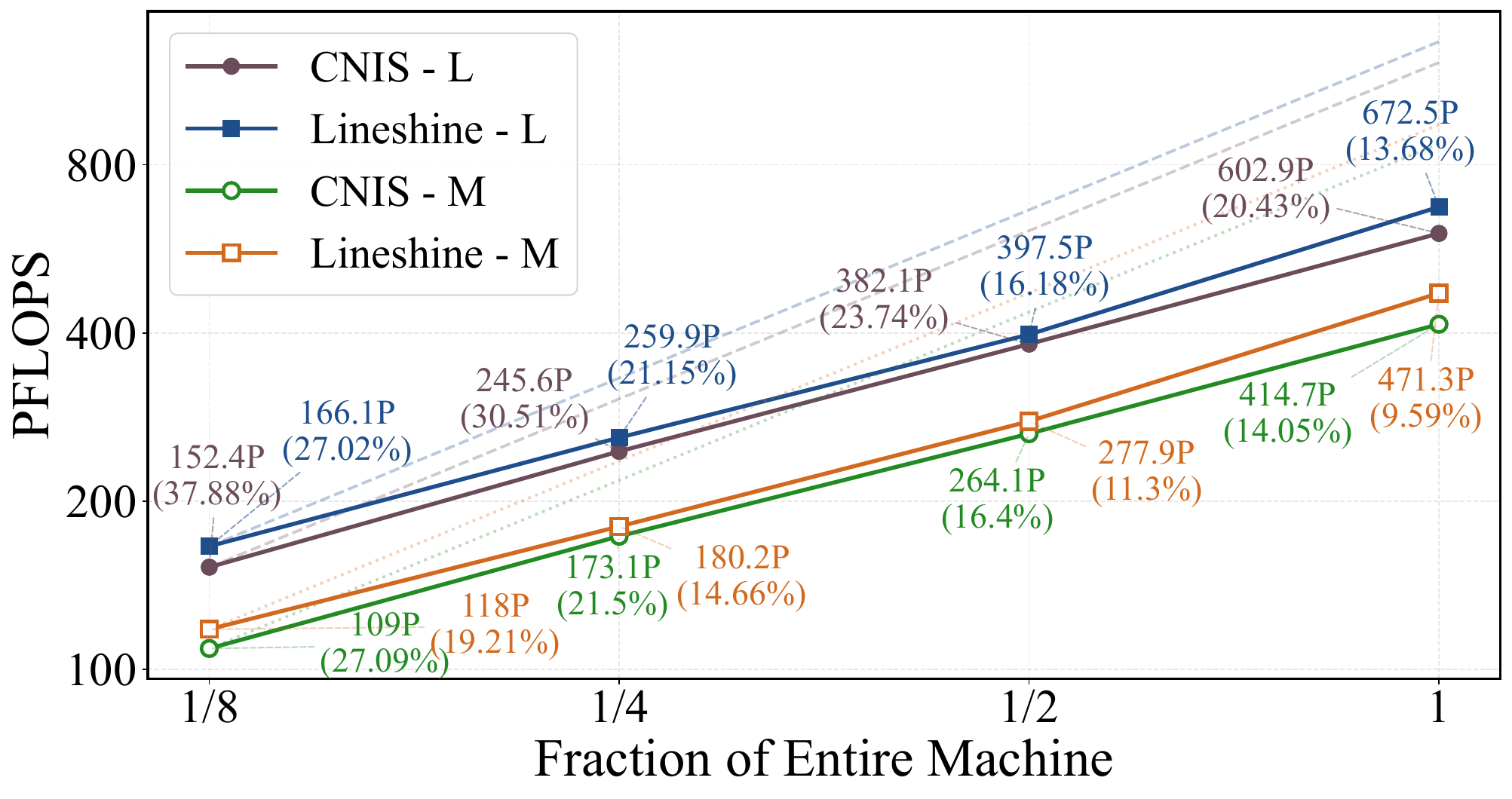}
    \caption{Strong scaling of MatRIS-MoE training on CNIS and LineShine. MatRIS-MoE (L) (solid markers) and MatRIS-MoE (M) (hollow markers) are scaled from 1/8 to full-machine scale with a fixed global problem size. The corresponding training throughput in PFLOPS and relative parallel efficiency are presented. Both models maintain over 50\% relative efficiency at full-machine scale.}
    \label{fig:strong_scaling}
\end{figure}

We evaluate the strong scaling of MatRIS-MoE training from 1/8 to full-machine scale while keeping the global batch size fixed for $1000$ iterations.
This reduces the workload per FS-3D unit through atomic-graph partitioning, decreasing the number of atoms per FS-3D unit from 128 to 64, 32, and 16 for MatRIS-MoE (M), and from 64 to 32, 16, and 8 for MatRIS-MoE (L).
As shown in Fig.~\ref{fig:strong_scaling}, both models scale up to full-machine scale.
\subsection{Strong Scaling}
\label{sec:strong_scaling}

Taking 1/8-scale performance as baseline, MatRIS-MoE (L) achieves parallel efficiencies of 53.93\% on CNIS and 50.60\% on LineShine at full scale, reaching peak performances of 602.9 PFLOPS and 672.5 PFLOPS, respectively. The average throughput scales from 0.80 million to 1.63 million edges/second on CNIS and from 0.90 million to 1.77 million edges/second on LineShine. For MatRIS-MoE (M), the efficiencies are 51.87\% and 49.9\%, with peak performances of 414.7 PFLOPS and 471.3 PFLOPS. The average throughput scales from  1.86 million to 3.92 million edges/second on CNIS and from 2.12 million to 4.24 million edges/second on LineShine.

{These consistent efficiencies across GPGPU-accelerated and many-core ARM platforms validate the portability of our Janus framework.}
The primary scaling bottlenecks arise from reduced GEMM sizes (due to fewer atoms per FS-3D unit), which lower arithmetic intensity and accelerator utilization, particularly for graph-based operators, as well as increased gradient All-Reduce communication overhead across more DP replicas, especially during second-order backpropagation.

\subsection{Weak Scaling}
\label{sec:weak_scaling}

In the weak scaling tests, the global batch size and the number of nodes are scaled proportionally from 1/8 to full-machine scale while maintaining the same intra-group configuration as in the strong-scaling experiments.
The number of atoms per FS-3D unit was kept constant at 128 for (M) and 64 for (L) across all scales.

Both models exhibit near-linear scaling. For MatRIS-MoE (L), \textbf{peak performance increases from 152.4 PFLOPS to 1,048.3 PFLOPS on CNIS (93.78\% efficiency) and reaches 1,200.2 PFLOPS (1.2 EFLOPS) on LineShine (90.3\% efficiency)}. For MatRIS-MoE (M), throughput scales from 109.0 PFLOPS to $750.9$ PFLOPS on CNIS (93.91\% efficiency) and reaches 861.6 PFLOPS on LineShine (91.3\% efficiency).

The only scaling overhead is gradient All-Reduce across expanding DP replica groups, which our optimized collectives handle efficiently. The sustained efficiency reaches 72.73\% on CNIS and 86.61\% on LineShine, demonstrating the effectiveness of our system–algorithm co-design.
%The sustained >90\% efficiency on both Exascale platforms demonstrates the effectiveness of our system-algorithm co-design.

\begin{figure}[!ht]
    \centering
    \includegraphics[width=\linewidth]{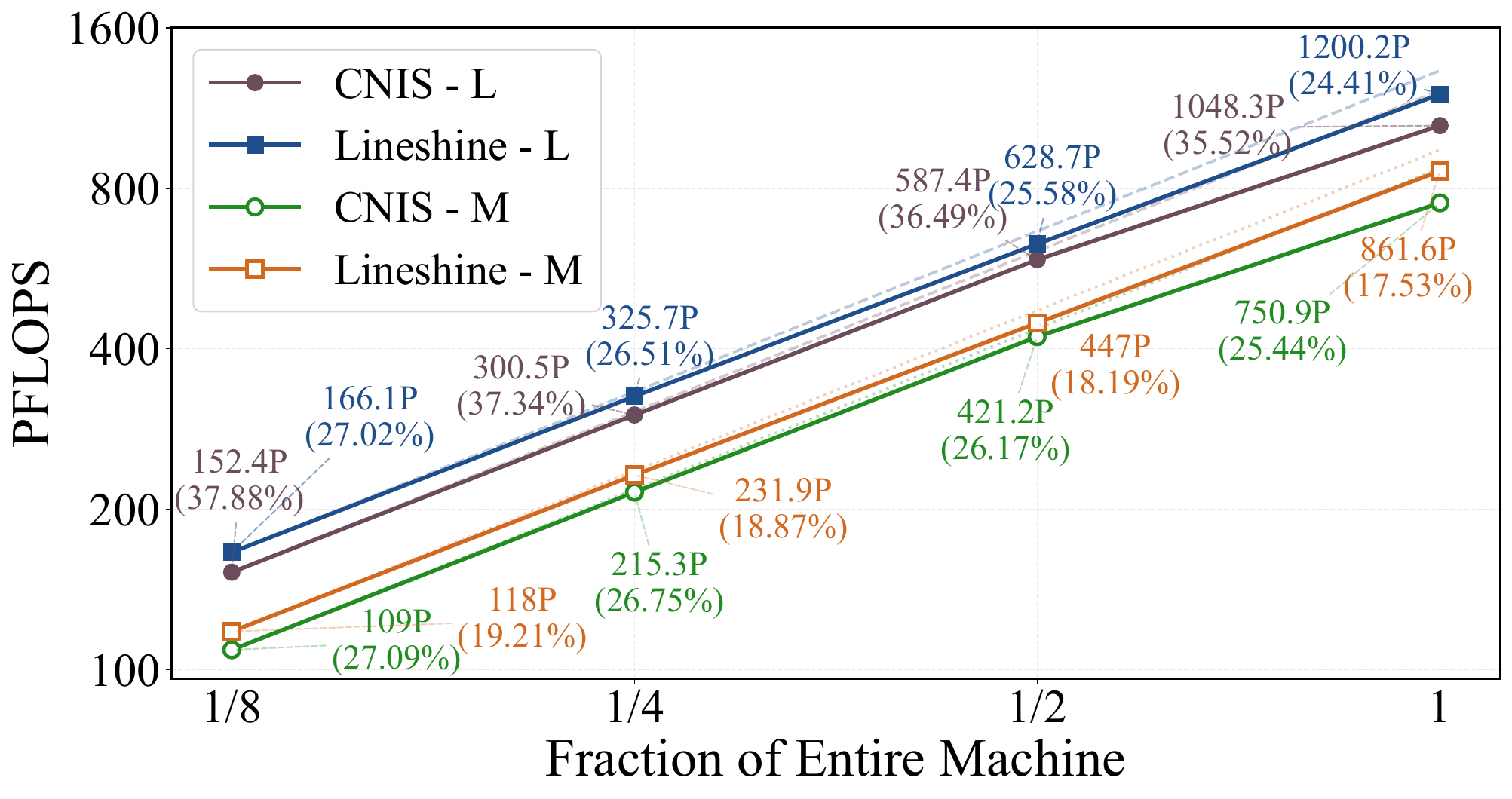}
    \caption{Weak scaling of MatRIS-MoE training on CNIS and LineShine. The problem size is increased proportionally with the machine scale from 1/8 to full-machine scale. The corresponding training throughput in PFLOPS is shown. Performance scales near-linearly, reaching 1.05 EFLOPS on CNIS and 1.20 EFLOPS on LineShine for MatRIS-MoE (L), with over 90\% parallel efficiency on both systems.}
    \label{fig:weak_scaling}
\end{figure}

\subsection{Sustained Performance}

While the peak performance in Sec.~\ref{sec:weak_scaling} represents steady-state throughput, sustained performance accounts for the end-to-end wall-clock time, encompassing overheads such as initialization and I/O. On CNIS, setup requires 591.3 seconds for RCCL initialization and 472.2 seconds for first-step warm-up. On LineShine, MPI initialization takes 923.5 seconds due to the larger node count, and Python module imports trigger additional HBM and PyTorch runtime initialization (531.8 seconds).

Measured over $1,000$ training steps on the full machine with the MatRIS-MoE (L), sustained performance reaches 762.3 PFLOPS (25.84\% of peak) on CNIS and 1,033.3 PFLOPS (21.02\% of peak) on LineShine. The gap between sustained and peak performance arises primarily from initialization overheads, which become increasingly pronounced at Exascale scale.

\section{Implications}
%\REQ{implications for future systems and applications (1 p max)}

\subsection{Implication for Applications}

\begin{figure}[htbp]
    \centering
    \includegraphics[width=\linewidth]{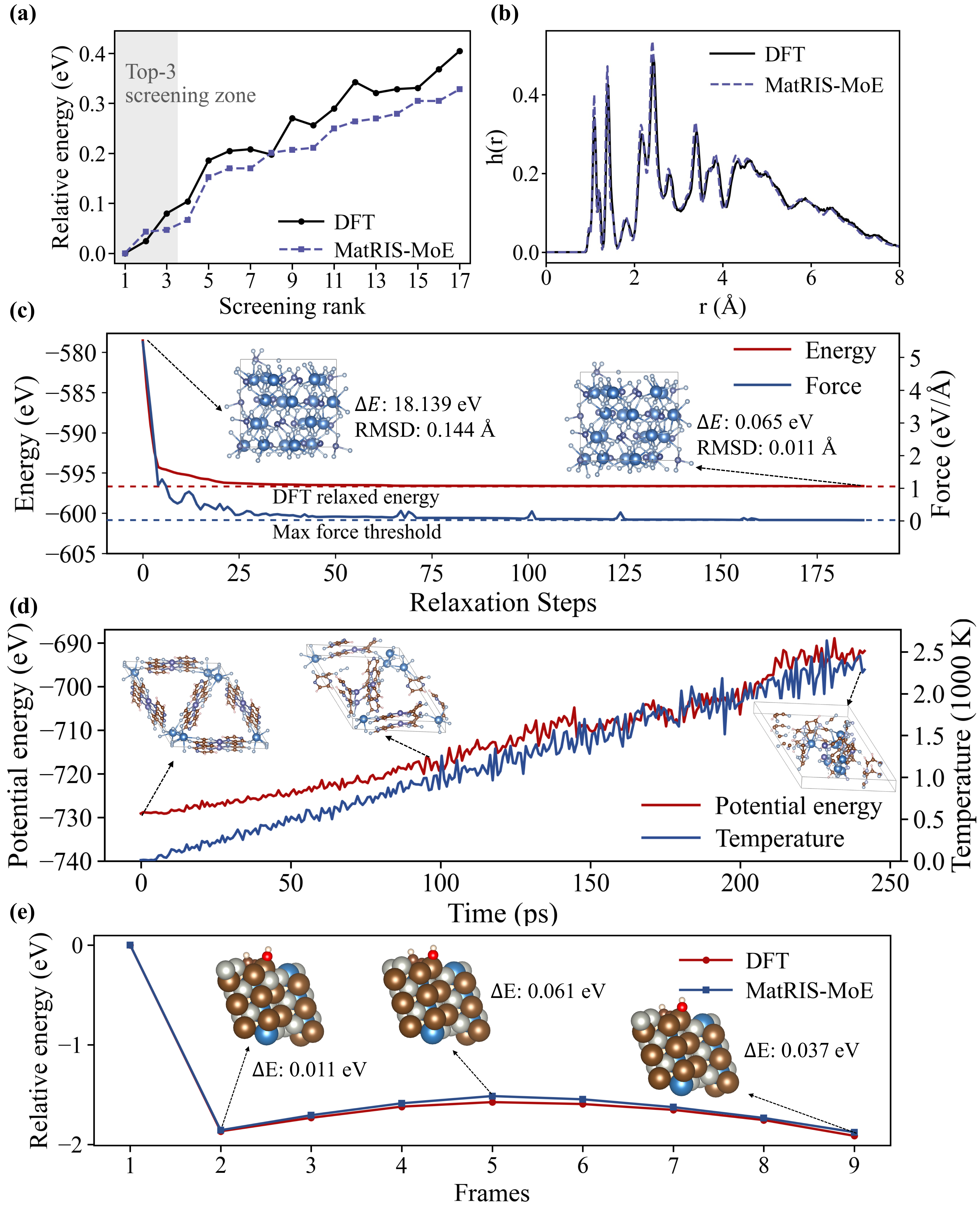}
    \caption{Applicability of MatRIS-MoE across (a) energy ranking, (b) molecular distribution functions, (c) structural relaxation, (d) molecular dynamics, and (e) catalytic reaction profiling.
    }
    \label{fig:Implications}
\end{figure}

Figure~\ref{fig:Implications} highlights the broad applicability and robustness of MatRIS-MoE across representative molecular, catalytic, and materials systems. Across a range of downstream tasks, the model remains in close agreement with DFT. 
In practical polymorph energy-ranking tasks~\cite{energy_ranking}, MatRIS-MoE predictions are very close to the DFT references (Fig.~\ref{fig:Implications}(a)). MatRIS-MoE can accurately reproduce the radial distribution functions of an Aspirin configuration in rMD17 (Fig.~\ref{fig:Implications}(b)). In the structural relaxation~\cite{matbench} of Ca$_{20}$O$_{60}$Se$_{20}$ (Fig.~\ref{fig:Implications}(c)), MatRIS-MoE efficiently drives the system toward a stable local minimum, with both the energy difference ($\Delta E$) and Root Mean Square Displacement (RMSD) approaching the DFT-relaxed reference. Continuous NVT molecular dynamics simulations~\cite{High_temper_phase_transition} of a complex metal-organic framework (Fig.~\ref{fig:Implications}(d)) further show that the model preserves structural integrity and thermodynamic stability as the temperature increases. 
MatRIS-MoE can also capture the relative energy profile along a representative catalytic reaction pathway~\cite{MD17_gas_phase} (Fig.~\ref{fig:Implications}(e)). 
Beyond the representative examples shown in Fig.~\ref{fig:Implications}, MatRIS-MoE can also support a wide range of practical applications, such as the prediction of mechanical properties, phonons, and phase diagrams.

MatRIS-MoE bridges quantum-level fidelity with enhanced representational power for heterogeneous, multi-component, and multi-task systems, and therefore serves as a powerful and versatile simulation engine for critical scientific challenges, including ion transport in solid-state electrolytes, active-site dynamics in single-atom catalysis, and the thermodynamic stability of complex doping configurations. Moreover, compared to traditional DFT calculations, MatRIS-MoE is two orders of magnitude faster.
%MatRIS-MoE learns intrinsic representations of atomic species and can distinguish the behavior of the same element across different chemical environments. 
Due to its high out-of-the-box predictive accuracy and strong generalization, MatRIS-MoE has the potential to further expand the horizon of MD simulations with \textit{ab initio} accuracy.
%replace DFT software while offering orders-of-magnitude speed advantages over DFT.
\subsection{Implication for HPC Systems}

This paper establishes billion-parameter universal MLIP training as a new Exascale AI-for-Science workload, and demonstrates that only through model–framework–system co-design can such workloads achieve efficient, end-to-end training at Exascale across many-core Exascale platforms. As the volume and diversity of \textit{ab initio} atomistic datasets continue to grow, pushing the boundaries of universal MLIPs will inevitably require scaling models beyond the 10-billion-parameter regime explored in this work. 

This trajectory, however, indicates that substantial future efforts must be directed toward extreme-scale system optimization. The effective utilization of dense matrix-multiply units (e.g., Tensor Cores and SME) will remain the cornerstone of arithmetic efficiency. Simultaneously, aggressively mitigating data movement, reducing memory footprints during second-order differentiation, and compressing inter-node communication overheads are imperative to sustain parallel efficiency. Conquering these system bottlenecks will be the definitive key to successful model scaling in the future.

Furthermore, the rapid evolution of AI4S signals a fundamental computing paradigm shift: AI workloads and traditional HPC simulation workflows are becoming equally important. With their full-precision capabilities (FP64/FP32/FP16), massive high-bandwidth memory, and tightly coupled interconnects, Exascale supercomputers are unequivocally the ideal platforms for the next-generation workloads. To fully harness these architectures, the community must develop more flexible and extensible hybrid programming frameworks that seamlessly integrate conventional scientific computing tools (e.g., molecular dynamics engines and DFT solvers) with large-scale distributed neural architectures.

Ultimately, as atomic foundation models continue to scale in both capacity and efficiency, their enhanced representational power will unlock the ability to tackle increasingly complex and high-impact scientific problems. By accurately capturing complex multi-body interactions across the periodic table, large-scale uMLIPs will bridge the long-standing gap between theoretical simulation and practical application, catalyzing transformative breakthroughs in clean energy, advanced materials, and next-generation pharmaceuticals.

\section*{Acknowledgment}

Numerical calculations are performed on the LineShine and the China New-generation Intelligent Supercomputer (CNIS) supercomputers. We thank NSCC-SZ for technical support throughout this project. This work was partially supported by the National  Science Foundation of China(92270206, T2125013, 62372435).

% \balance
\vspace{5mm}

%\bibliographystyle{./IEEEtran}
%\bibliography{./IEEEabrv,./Reference}
%\clearpage
\bibliographystyle{IEEEtran}
\bibliography{Reference}

\end{document}

%% file: 3-model.tex
\subsection{Algorithmic Innovation}\label{subsec: model inno}
\begin{figure*}[!t] 
    \centering
    \includegraphics[width=\textwidth]{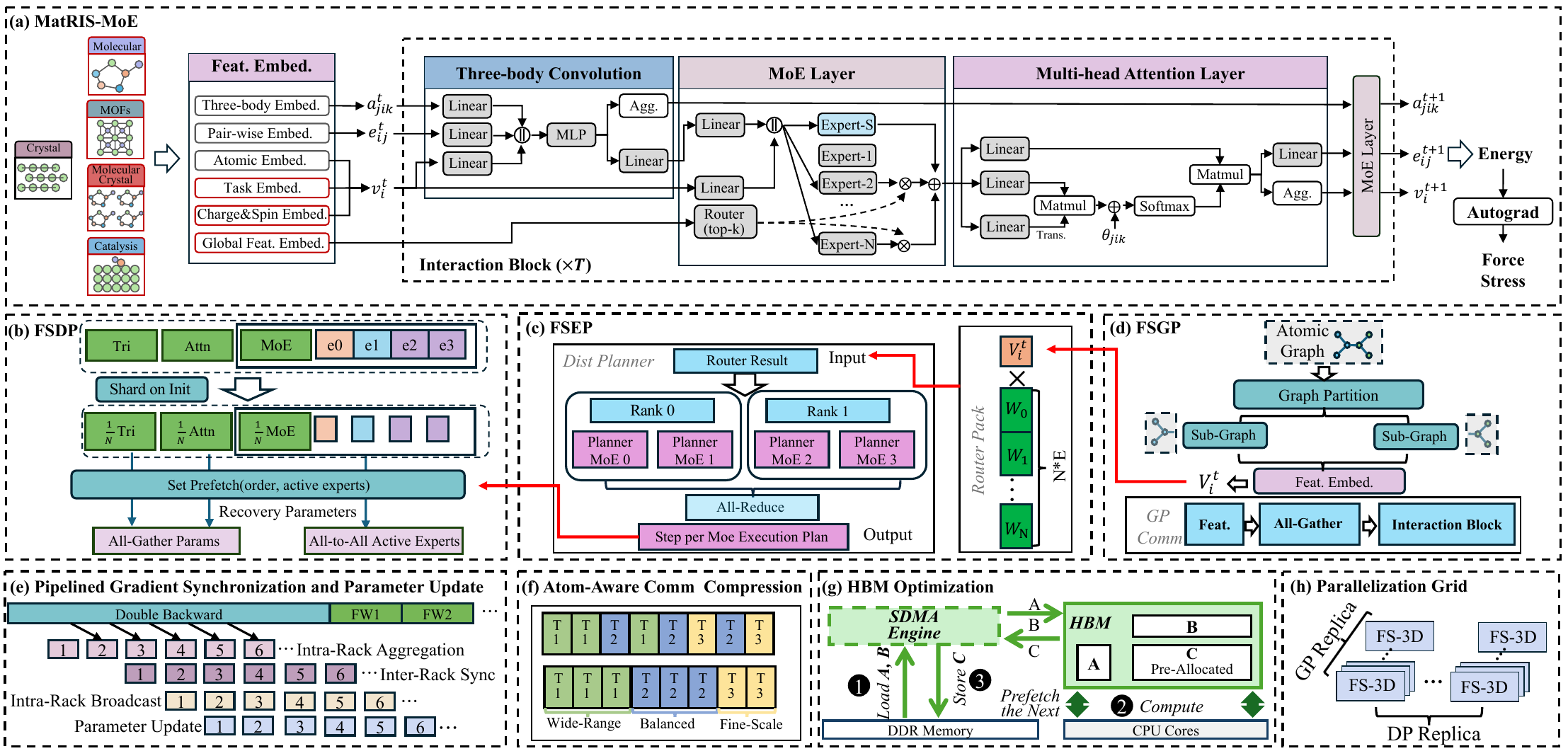}
    \caption{ Overview of our work.  (a) Model architecture.  (b)--(d) Framework-level optimizations, including FSDP, FSEP, and FSGP.  (e)--(g) Supercomputer-level optimizations for communication and memory efficiency. (h) Parallel strategy of DP and GP replica with FS-3D }
    \label{fig:model arch}
    \label{fig:overview}
\end{figure*}
% Siyu(04-15)
Extending the single-task MatRIS to a multi-task model that jointly supports isolated molecules, periodic crystals, catalytic systems, MOFs, and periodic molecules presents three core challenges: resolving dataset inconsistencies (e.g., varying DFT functionals), balancing massive model capacity with computational feasibility, and fully utilizing GPU hardware for modelling complex many-body interactions. 

We address these challenges through the following architectural and training innovations.

\subsubsection{\textit{Task-aware feature embedding}} 
To align the systems that are calculated under different density functionals (e.g., PBE, R2SCAN, $\omega$B97M) within a shared representation space, MatRIS-MoE augments the feature embedding with three additional components: a task embedding that injects dataset embeddings, charge and spin embeddings, and a global feature embedding to distinguish element composition. The computational cost of the embedding module remains strictly linear in the graph size. 

\subsubsection{\textit{Sparse mixture-of-experts (MoE) routing}}
MatRIS-MoE inserts sparse expert modules both before and after the self-attention layer. A message-update MoE to specialize in message construction and a feature-update MoE to specialize in post-attention refinement. In each MoE layer, routing is determined by element types. Each element type activates its own top-$K$ experts. Because element types are time-independent, the activated expert set remains stable during simulation, thereby maintaining a continuous and smooth PES. 
In multi-task modeling, this sparse element-wise routing allows the model to handle a wider variety of tasks. Different experts can specialize in the chemistry associated with different elements, thereby improving expressiveness compared to the dense parameterization setting. 
During training, expert parallelism enables efficient distributed execution across large-scale clusters.
Once the model is trained, during inference, only the top-$K$ routed experts for each element are activated, so the effective parameter number and FLOPs are much smaller than those of the full MatRIS-MoE model. When the number of activated experts is $K$, and the hidden width is $d$, the total parameter number of one MoE layer grows as $O(K \cdot d^2)$.
% \textcolor{red}{the activated expert is determined by atom types}

\subsubsection{\textit{Multi-head self-attention}}
MatRIS employs separable attention to reduce the complexity of dense pairwise attention from $O(N^2)$ to $O(N)$ by decoupling source- and target-aware aggregation. However, multi-task learning across heterogeneous chemical domains requires greater expressive power to capture complex, task-dependent many-body interactions. To address this limitation, MatRIS-MoE replaces the original separable attention with multi-head self-attention. Under a fixed cutoff radius, the computational complexity of multi-head self-attention is $O(m \cdot d \sum_i |\mathcal{N}(i)|^2)$, where $m$ is the number of attention heads, $d$ is the hidden dimension, and $|\mathcal{N}(i)|$ denotes the number of neighbors of atom $i$. Although this quadratic scaling with respect to the number of neighbors increases the algorithmic cost, the compute-intensive multi-head self-attention maps more efficiently onto dense matrix operations, enabling superior utilization of modern GPU hardware for both training and inference.
% The multi-head self-attention offers much higher arithmetic intensity.

\subsubsection{\textit{Multi-task training strategies}}
MatRIS-MoE adopts a conservative training manner, where forces and virial stresses are obtained strictly by automatic differentiation of the predicted energy.
To enhance the optimization stability in the multi-task setting, MatRIS-MoE adopts a multi-task robust training loss:
%\begin{equation}
    $\mathcal{L}_{\mathrm{robust}} = \frac{1}{N} \sum_{i=1}^{N} \mathcal{W}(z_i, \tau)^2 \cdot L_i$
%\end{equation}

Here, $N$ denotes the number of structures in a batch, and $L_i$ is the base loss of the $i$-th sample (e.g., L1 or Huber loss), defined as a weighted combination of errors in system energy ($e$), atomic forces ($f$), stress ($s$), and magnetic moments ($m$). For each forward pass, we compute the mean $\mu$ and standard deviation $\sigma$ of the batch losses and use them to form the standardized score $z_i=(L_i-\mu)/\sigma$. The weighting function $\mathcal{W}(z_i, \tau)$ is a smooth monotone soft-thresholding function. Samples in the normal range ($z_i \le \tau$, with $\tau=2.0$ in this work) retain weights close to 1, whereas extreme outliers are progressively down-weighted toward 0. To avoid interference among heterogeneous tasks, the loss statistics used to evaluate $z_i$ are computed independently for each task.

%% file: 012-system.tex
\subsection{Janus: the uMLIP Training Framework}
Janus is designed to support agile development of MatRIS-MoE while delivering high efficiency at exascale, with two primary goals: minimizing code intrusion to preserve rapid model iteration, and supporting sharded execution across both double-backward training and MoE layers.
Janus adopts FS-3D (Fully Sharded 3 Dimensions) as its basic execution unit, and replicates this unit along the data-parallel (DP) and graph-parallel (GP) dimensions (see Fig.~\ref{fig:overview}(h)).
Within each DP-replica, the global batch is first divided into mini-batches. Each mini-batch is then partitioned across GP-replicas, and the resulting subgraphs are further partitioned along the intra-FS-3D graph dimension. 
To the best of our knowledge, Janus is the first hybrid-parallel training framework for uMLIPs.
The key contributions are as follows.

\subsubsection{{FS-3D, a unified hybrid-parallel runtime for training MatRIS-MoE}}
FS-3D integrates FSDP (fully sharded data parallelism), FSGP (fully sharded graph parallelism), and FSEP (fully sharded expert parallelism) into a unified execution unit for large-scale training.
As illustrated in Fig.~\ref{fig:overview}(b)-(d), FS-3D initializes sharded interaction blocks with FSDP, partitions the atomic graph across ranks with FSGP, and then executes sharded MoE layers with expert parallelism (FSEP).
Specifically, 
\ding{202} FSDP reduces static memory usage, including model parameters, gradients, and optimizer states. 
Each GPU stores only a local parameter shard. Before computation, FS-3D restores the full parameters of non-MoE layers through all-gather, whereas MoE expert parameters are restored through all-to-all communication.
After computation, the restored parameters are re-sharded to recover the memory savings of sharded training.
\ding{203} FSGP reduces activation memory usage for large graph batches. Built on GP~\cite{GP_sriram2022towards}, it partitions a large atomic graph (batch) across multiple GPUs, so that each GPU stores and computes only its local partition.
\ding{204} FSEP improves load balance across devices. 
Inspired by LAER-MoE~\cite{fsep_LAER-MoE}, FSEP partitions expert parameters across devices and restores only the active experts via all-to-all communication.
We further design a just-in-time sparse expert planning mechanism to achieve finer-grained load balancing and eliminate redundant communication.
We also reorder collective communication operators to ensure both correctness and performance (see Fig.~\ref{fig:fs3d-timeline}).

\subsubsection{{Just-in-time sparse expert planning}}
Building on FS-3D, we further address two core inefficiencies in MoE execution:
(i) each training step activates only a small subset of experts, yet conventional FSEP still keeps many unused expert parameters; 
(ii) the active expert set and its token load can vary sharply across steps in heterogeneous material workloads, making prior approaches (e.g., static expert-to-rank binding, and history-based re-layout) brittle.
Because MatRIS-MoE routes experts based on global atomic embeddings and neighborhood information, all MoE layers can be routed before entering the interaction blocks.
To leverage this feature, we adopt a replica-free expert execution approach driven by just-in-time (JIT) planning (see Fig.~\ref{fig:overview}(c)).
\ding{202} \textit{Plan Generation.} 
At the beginning of each training step, we perform batched routing for all MoE layers and collect per-layer token counts.
On each device, we run a local planner based on token counts, which inserts experts into a heap by load, repeatedly pops the heaviest expert, and deterministically assigns it to the least-loaded rank.
The global plan, containing the expert prefetch order and token dispatch splits required by the FS-3D runtime, is derived by merging the local plans. 
Therefore, active experts can be placed with both balanced load and a reproducible owner layout.
\ding{203} \textit{Expert Sparsification.} 
Based on this plan, FS-3D avoids all-gathering the full expert set in the forward pass.
Instead, only active experts are materialized through parameter all-to-all communication along the shard dimension, and each active expert is recovered only on its owner rank.
\ding{204} \textit{Neighbor Feature Dispatch.}
FS-3D then dispatches the routed neighbor-atom features $\mathrm{atomic}_j$ to the ranks that host the corresponding active experts and gathers the expert outputs through all-to-all communication.
\ding{205} \textit{Gradient Combination.}
In the backward pass, sparse expert gradients are returned through all-to-all gradient exchange along the shard dimension, while replica synchronization is restricted to the union of active experts in the global batch. 
To preserve consistency under asymmetric expert activation, ranks explicitly zero-fill missing gradients before sparse synchronization along the replica dimension.

\subsubsection{{Runtime for multi-phase double-backward execution}}
\begin{figure}
    \centering
    \includegraphics[width=\linewidth]{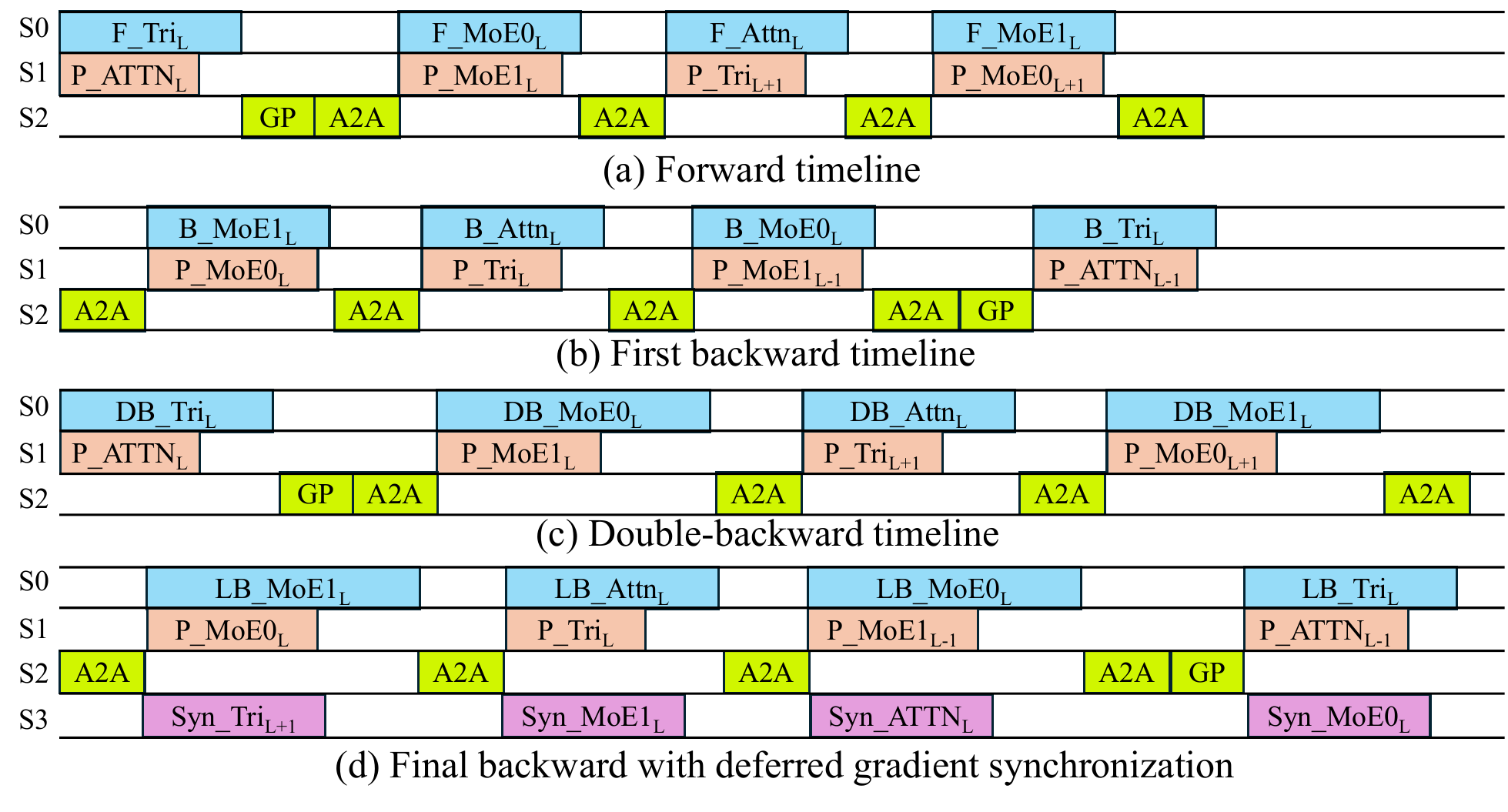}
    \caption{Execution timeline of our framework in MatRIS-MoE. Each interaction block contains Tri, MoE0, Attn, and MoE1. Blue blocks denote operator computation, pink blocks denote parameter prefetch, green blocks denote GP communication or token-routing All-to-All (A2A), and purple blocks denote deferred gradient synchronization.}
    \label{fig:fs3d-timeline}
\end{figure}

For MatRIS-MoE, each iteration consists of four phases: forward for energy computation, first backward for force computation, double backward for force-loss propagation, and final backward for energy-loss propagation.
FS-3D restores parameters before each phase and re-shards them after use through operator-specific communication. It also records the execution order in the forward phase and reuses it in later phases to enable prefetch and overlap.
We introduce a double-backward-aware lifecycle that restores parameters on demand and defers gradient synchronization until the final backward phase.
Fig.~\ref{fig:fs3d-timeline} shows the execution timeline of an FS-3D unit.

\subsubsection{{Load balancing}}
To mitigate load imbalance caused by variable atomic graph sizes, we adopt a deterministic greedy batching strategy: samples are inserted into a max-heap by atom count, and iteratively assigned to the currently least-loaded mini-batch.
This load balancing is performed only within each global batch and does not change the parameter update granularity or training semantics.

\subsection{Training Optimizations on CNIS Supercomputer} %\wy{:delete supercomputer here}}

We train MatRIS-MoE on China's New-generation Intelligent Supercomputer (CNIS), which is equipped with SIMT-based GPGPU accelerators with 64 GB of HBM2e memory each.
Each node contains 8 GPGPUs, and the training leverages the Janus framework.
% The software stack is based on HIP, RCCL, rocBLAS, and PyTorch. 

\subsubsection{{Pipelined gradient synchronization and parameter update}}
\label{sec:pipelined_comm}
To reduce the parameter update overhead of synchronous training, we design a pipelined scheduling mechanism across multiple gradient buckets, coordinating the hierarchical communication and parameter updates. 
As shown in Fig.~\ref{fig:overview}(e), for each bucket, we preserve the dependency chain of intra-rack aggregation $\to$ inter-rack synchronization $\to$ intra-rack broadcast $\to$ parameter update.
We leverage the fact that current optimizers introduce no data dependency across gradient buckets.
Therefore, our scheduler parallelizes the execution of these operations across different buckets and effectively fills the pipeline while minimizing bubbles.
In addition, our scheduler ensures that computation for the next iteration's forward pass corresponding to bucket$_i$ can begin immediately after the parameter update for bucket$_i$ is completed.

\subsubsection{{Atom-type-aware communication compression}}
All-to-all communication in MoE layers scales with token count and feature dimension.
At the scale of MatRIS-MoE, a single all-to-all incurs a large communication volume per layer.
We reduce the communication cost using an atom-type-aware FP16 compression scheme (see Fig.~\ref{fig:overview}(f)).
After routing, tokens in the local dispatch buffer are grouped by atom type, quantized with per-type scaling factors, transferred in FP16, and dequantized after communication, before restoring the original order.
We exploit the fact that tokens from the same chemical element have similar activation ranges, achieving lower quantization error than coarse-grained per-tensor scaling while maintaining convergence.

\subsubsection{{Kernel engineering for critical operators}}
MatRIS-MoE operates on atomic graphs with irregular neighbor structures, making several kernels severely memory-bound.
We optimize four major performance-critical kernels: neighbor gather, edge aggregation, multi-head attention, and MoE dispatch/combine.
The optimizations include fused indexing for coalesced memory access, GEMM-based reformulation of reductions, batched GEMM invocation, and contention-aware scatter.
These kernel-level optimizations substantially improve single-accelerator throughput and remain effective at scale in distributed training.

\subsection{Training Optimizations on LineShine Supercomputer}
In this section, we describe the training optimizations on the LineShine platform. The system has an asymmetric memory topology: each LX2 CPU socket contains two compute dies, with each die integrating four NUMA domains (38 ARMv9 cores and 4 GB of HBM per domain) and a dedicated SDMA engine.

\subsubsection{{Async MPI via software-defined streams}}
FS-3D orchestrates the ordering and dependencies of collective communications, and relies on CUDA streams for asynchronous communication and computation overlap.
Given that PyTorch's CPU backend lacks the asynchronous CUDA stream semantics available on GPUs, we address this limitation by developing a software-defined asynchronous MPI runtime.
Recognizing that the LX2 is a core-rich processor architecture, we physically partition CPU resources.
We partition CPU resources as follows:
\ding{202} computation cores: 32 cores are pinned for OpenMP computation, which is sufficient to saturate memory bandwidth;
\ding{203} communication cores: 5 dedicated cores are reserved exclusively for the MPI communication thread pool;
and \ding{204} 1 core reserved for the operating system. 
By offloading collective operations to communication cores via \textit{taskflow} task graphs, we successfully emulate CUDA-like stream behavior, thereby avoiding explicit inter-NUMA SDMA calls and keeping the computation cores unblocked.
This enables robust overlap between communication and computation, allowing Janus to capitalize on the advanced scheduling optimizations described above.

\subsubsection{{HBM memory optimization via SDMA engine}}
To mitigate DDR contention, we design an explicit memory hierarchy management strategy that utilizes the on-package SDMA engines.
We customize the HBM data movement directly at the memory-bound operator level rather than relying on compiler-driven graph optimizations. 
We use the optimized GEMM operator as an example.
As illustrated in Fig.~\ref{fig:overview}(g), it contains three phases:
\ding{202} upon invocation, the operator synchronously dispatches SDMA instructions to bulk-transfer both the input activations $A$ and weights $B$ from DDR to HBM buffers; 
\ding{203} the GEMM computation $C {=} AB$ is performed using pre-allocated HBM buffers; 
\ding{204} once the computation is completed, SDMA is invoked again to write the result $C$ back to DDR memory.
We implement a double-buffered SDMA pipeline: while the CPU computes layer $k$ in HBM, the SDMA engine asynchronously prefetches layer $k{+}1$ from DDR, effectively hiding the memory transfer latency.
By offloading large transfers (e.g., tensors exceeding 1\,MB) to the SDMA, we achieve up to a 1.4$\times$ bandwidth improvement over CPU-driven transfers under full core utilization, while simultaneously freeing valuable CPU cycles for subsequent instruction decoding.